\documentclass{article} 
\usepackage[english]{babel}
\usepackage[utf8]{inputenc}
\usepackage{amsmath}
\usepackage{graphicx}
\usepackage{wrapfig}
\usepackage{amsthm}
\usepackage{amsfonts}
\usepackage{amssymb}
\usepackage[usenames,dvipsnames]{xcolor}
\usepackage{graphicx}
\usepackage[siunitx]{circuitikz}
\usepackage{tikz}
\usepackage{wrapfig}
\usepackage[colorinlistoftodos, color=orange!50]{todonotes}
\usepackage{hyperref}
\usepackage[numbers, square]{natbib}
\usepackage{fancybox}
\usepackage{epsfig}
\usepackage{soul}
\usepackage{forest}
\usepackage{setspace}
\usepackage{xcolor}
\hypersetup{
    colorlinks,
    linkcolor={red!50!black},
    citecolor={blue!50!black},
    urlcolor={blue!80!black}
}
\usepackage[framemethod=tikz]{mdframed}
\usepackage[shortlabels]{enumitem}
\usepackage[version=4]{mhchem}
\usepackage{multicol}
\usepackage{wrapfig,lipsum,booktabs}
\newcommand{\sys}[1]{\texttt{ATLASv2}\textsubscript{#1}}

\usepackage{changepage}

\title{{\vspace{-2cm}}
\rule{\linewidth}{0.5pt} \\[6pt] 
\huge ATLASv2: ATLAS Attack Engagements, Version 2 \\
\rule{\linewidth}{0.5pt}  \\[10pt]
}
\author{Andy Riddle, Kim Westfall, and Adam Bates\\University of Illinois at Urbana-Champaign}
\date{}

\begin{document}

\maketitle

\section{Introduction}
\sys{} is based on a previously generated dataset included in ATLAS: A Sequence-based Learning Approach for Attack Investigation [1]. The original ATLAS dataset is comprised of Windows Security Auditing system logs, Firefox logs, and DNS logs via WireShark. In \sys{}, we aim to enrich the ATLAS dataset with higher quality background noise and additional logging vantage points. This work replicates the ten attack scenarios described in ATLAS, but extends the logging to include Sysmon logs and events tracked through VMware Carbon Black Cloud.\\

The main contribution of \sys{} is to improve the quality of the benign system activity and the integration of the attack scenarios. Instead of relying on automated scripts to generate activity, we had two researchers use the victim machines as their primary work stations throughout the course of the engagement. This allowed us to capture system logs on actual user behavior. Additionally, the researchers conducted the attacks in a lab setup allowing the integration of the attack into the work flow of the victim user. This allows the \sys{} dataset to provide realistic system logs that mirror the system log activity generated in real-world attacks.\\

\begin{table}[h]
  \centering
  \begin{tabular}{@{}r l l l@{}}
      \textbf{Attack} & \textbf{CVE} & \textbf{Exploit}\\
    \hline
s1 & CVE-2015-5122 & Adobe Flash\\
s2 & CVE-2015-3105 & Adobe Flash\\
s3 & CVE-2017-11882 & Microsoft Word\\
s4 & CVE-2017-0199 & Microsoft Word\\
m1 & CVE-2015-5122 & Adobe Flash\\
m2 & CVE-2015-5119 & Adobe Flash\\
m3 & CVE-2015-3105 & Adobe Flash\\
m4 & CVE-2018-8174 &  Microsoft Word\\
m5 & CVE-2017-0199 &  Microsoft Word\\
m6 & CVE-2017-11882 &  Microsoft Word\\
    \hline
    
  \end{tabular}
  \caption{\label{attack:scenarios}\textbf{Attack Scenarios}.}
\end{table}

{\vspace{-.75cm}}
\section{Methods}
The \sys{} dataset was generated by two researchers, each on their own VM. The dataset covers a four day benign period followed by a fifth ``attack day" where the ten scenarios (listed in Table\ref{attack:scenarios}) were executed on those VMs.\\
\\
All logs included in this dataset were generated on two Windows 7 32-bit VMs which we refer to as host 1 (h1) and host 2 (h2). The Windows VMs were configured with Mozilla Firefox 52.0, Microsoft Office Professional Plus 2010, Adobe Flash Player 17.0.0.188 and 18.0.0.194 (varied by attack scenario), and included Carbon Black sensors v.3.8.0.627 with data-forwarders set to an AWS S3 bucket. A third machine running Kali Linux was used to execute the attack scenarios on the fifth day of the engagement.
\\{\vspace{-.05cm}}
\subsection{Machine Instrumentation}
The \sys{} machines were instrumented with several audit systems at the OS-level and application-level. The following audit telemetry streams were recorded for both hosts:
\begin{itemize}
    \item Microsoft Windows ETW Security Audit.
    \item Wireshark for DNS logging information.
    \item Carbon Black Cloud sensor version 3.8.0.627.
    \item Firefox application logs.
    \item SysMon System monitor for Windows.
\end{itemize}

\subsection{Benign Period}
System logging events cover a five-day period. The first four days of the engagement records two researchers using the Windows VMs as their primary workstations approximately 8-hours per day, simulating a normal work day. At the conclusion of each workday, the VMs are left running overnight. All system activity was generated manually by the two researchers assigned to the VMs. We refer to all events during these four days as the benign period.\\
\\
Over the course of the benign period, the two researchers used a variety of applications from their usual work behavior such as web browsing, attending Zoom meetings, working on projects using Microsoft Office applications, sending and receiving e-mail attachments, chatting on Discord, watching videos, and other normal user behavior. The user on the h1 machine also ran a SimpleHTTP server to simulate an internal web portal. This server is exploited in the multi-host attack scenarios to achieve lateral movement.
\subsection{Attack Day}
The fifth day begins with benign activity then transitions into execution of the ten attack scenarios. Each of the ten attacks is assigned to a time window of about 30 minutes to 2 hours.\\
\\{\vspace{-.8cm}}
\begin{table}[h]
  \centering
  \begin{tabular}{@{}r l c@{}}

      \textit{Attack} & \textit{Machine Running Times} & \textit{Initial Attack Step}\\
    \midrule
\textbf{s1} & \texttt{2022-07-19 13:12:00-13:40:00} & \texttt{13:26:00}\\
\textbf{s2} & \texttt{2022-07-19 13:45:00-14:20:00} & \texttt{14:07:00}\\
\textbf{s3} & \texttt{2022-07-19 14:20:00-15:05:00} & \texttt{15:36:00}\\
\textbf{s4} & \texttt{2022-07-20 00:31:00-01:00:00} & \texttt{00:52:00}\\
\textbf{m1} & \texttt{2022-07-19 16:00:00-17:50:00} & \texttt{17:20:00}\\
\textbf{m2} & \texttt{2022-07-19 19:32:00-20:02:00} & \texttt{19:44:00}\\
\textbf{m3} & \texttt{2022-07-19 20:06:00-20:40:00} & \texttt{20:28:00}\\
\textbf{m4} & \texttt{2022-07-19 22:31:00-23:04:00} & \texttt{22:49:00}\\
\textbf{m5} & \texttt{2022-07-19 23:16:00-23:46:00} & \texttt{23:38:00}\\
\textbf{m6} & \texttt{2022-07-19 23:54:00-00:27:00} & \texttt{00:10:00}\\
    \bottomrule
    
  \end{tabular}
  \caption{\label{attack:timestamps}\textbf{Attack Timestamps}. UTC Timestamps for attack windows and initial attack step.}
\end{table}

{\vspace{-.05cm}}
Single-host attacks (s1-s4) involve only the Host 1 machine. For each single host attack, the user of h1 continues their normal work-flow while the attacker executes steps 1-5 (Table~\ref{attack:steps}) on the Kali Linux machine. Attacks are initiated at a random time within the given attack window.\\
\begin{table}[h]
  \centering
  \begin{tabular}{@{}l l@{}}

\footnotesize
\textbf{Step1} & \footnotesize{Attacker sends a phishing email.}\\
\footnotesize
\textbf{Step2} & \footnotesize{Victim visits webpage or downloads attachment based on specific attack CVE.}\\
\footnotesize
\textbf{Step3} & \footnotesize{Attacker gains access to a Meterpreter shell.}\\
\footnotesize
\textbf{Step4} & \footnotesize{Attacker drops an additional payload onto h1’s machine.}\\
\footnotesize
\textbf{Step5} & \footnotesize{Payload collects PDFs and opens an https connection to exfiltrate the files.}\\
\footnotesize
\textbf{Step6} & \footnotesize{Attacker overwrites h1’s index webpage with phishing page.}\\
\footnotesize
\textbf{Step7} & \footnotesize{H2 visits h1’s webpage and downloads payload.}\\
\footnotesize
\textbf{Step8} & \footnotesize{Payload collects PDFs and opens an https connection to exfiltrate the files.}\\

  \end{tabular}
  \caption{\label{attack:steps}\textbf{Attack Steps}.}
\end{table}

Multi-host attacks (m1-m6) begin with Host 1, then overwrite the index page on Host 1's SimpleHTTP server to then infect Host 2. In multi-host scenarios, both users generate benign data while the attacker performs attack steps 1-8.\\
\begin{wrapfigure}[15]{r}{0.35\textwidth}
\vspace{25pt}
\begin{centering}
\includegraphics[width=4cm]{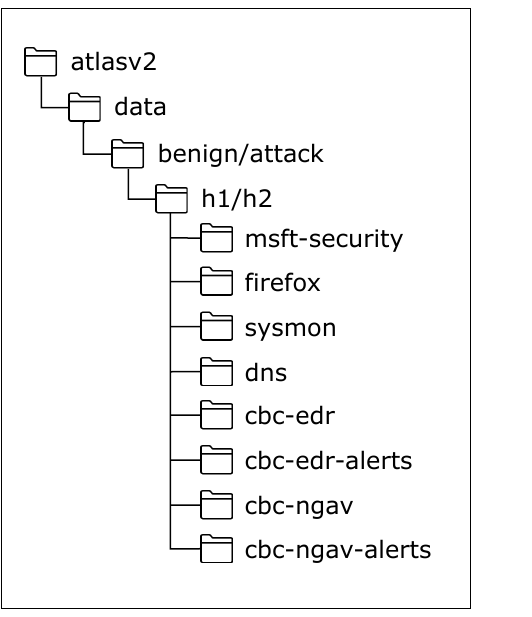}\caption{Directory Structure}\label{directory}
\end{centering}
\vspace{-175pt}
\end{wrapfigure}

{\vspace{-.05cm}}
\subsection{Data Collection and Organization}
To reduce file size, Microsoft Windows Security Auditing, Microsoft Sysmon, Firefox application system logging, and
DNS logs from Wireshark were moved off the VMs through a shared folder on the host machine at the conclusion of each working day. All Carbon Black Cloud data was sent to an AWS S3 bucket using data forwarding.\\
\\
The five day \sys{} engagement generated 154 GB of data. This includes raw logs from the five logging frameworks in the instrumentation on each machine. Logs are separated by host machine and organized by logging framework (Figure 1). In addition, Carbon Black Cloud logs are split by event type.\\

\section{Malicious Activity}

We operate under the assumption that all activity logged during the first four days is benign. Malicious activity begins with the initial attack step for each attack scenario. The timestamp of each initial attack step is recorded in Table~\ref{attack:timestamps}.\\
\\
The attacks employ one of two initial activities: either clicking on a malicious link to (attack) Adobe Flash, or downloading a malicious Word file. The known attack entities are {\texttt{ortrta.net}, 
\texttt{payload.exe}, 
\texttt{10.193.66.115:9999}, 
\texttt{10.193.66.115:8443}, 
\texttt{s3take2.zip}, 
\texttt{s4-at-night.zip}, 
\texttt{m4.zip}, 
\texttt{m5-2.zip}, and
\texttt{m6.zip}.\\
\\
 Ground truth labels for Microsoft Windows Security Auditing can be found at:\\
 \texttt{/atlasv2/data/attack/h*/msft-security/groundtruth/}.\\
 Labels are provided per-attack. Each file contanis a list of malicious events' "EventRecordID"  corresponding to the matching host-attack \texttt{msft-security-*-*.xml} file. 

\section{Limitations of the \sys{} Dataset}
 The \sys{} dataset attacks all share a common set of attack steps. While the initial point of intrusion utilizes various vulnerabilities and exposure methods, the {payload.exe} method is {the same across all attacks}. Additionally, the attack vectors are all either a malicious website or a malicious Microsoft Word file. This level of similarity between the attack steps leads to similarity in the logging data collected during each attack. Therefore, this dataset is not a good candidate for multi-class supervised learning.

 \subsection{Threats to Validity}
 Each attack scenario was executed on a different snapshot of the Windows 7 Virtual Machines. This leads to a startup footprint at the beginning of each attack scenario that creates a footprint at the start of each attack log. When using \sys{} for anomaly detection, this recurring sequence should be considered.
 
 \section{Availability}
 
 The ATLASv2 dataset is available for public release at \url{https://bitbucket.org/sts-lab/atlasv2}.
 Process labels for the attacker-controlled processes are available as part of the REAPr ground truth label set \url{https://bitbucket.org/sts-lab/reapr-ground-truth}.

\section*{Acknowledgements}

We would like to thank Abdulellah Alsaheel for his time and generosity in discussing 
  the original ATLAS dataset and sharing his materials with us.

{\vspace{.6cm}}

\end{document}